\documentclass[twocolumn,pra,showpacs,amsmath,amssymb]{revtex4-1}
\usepackage{graphicx}
\usepackage{bm}
\usepackage{epstopdf}
\usepackage{epsfig}
\usepackage{dsfont}
\begin{document}

\title{Space-time symmetric extension of non-relativistic quantum mechanics}
\author{Eduardo O. Dias}
\email[]{corresponding author: eduardodias@df.ufpe.br}
\author{Fernando Parisio}
\email[]{parisio@df.ufpe.br}
\affiliation{Departamento de
F\'{\i}sica, Universidade Federal de Pernambuco, Recife, Pernambuco
50670-901, Brazil}

\begin{abstract}
In quantum theory we refer to the probability of finding a particle
between positions $x$ and $x+dx$ at the instant $t$, although we
have no capacity of predicting exactly when the detection occurs. In
this work, first we present an extended non-relativistic quantum
formalism where space and time play equivalent roles. It leads to
the probability of finding a particle between $x$ and $x+dx$ during
[$t$,$t+dt$]. Then, we find a Schr\"odinger-like equation for a
``mirror'' wave function $\phi(t,x)$ associated with the probability
of measuring the system between $t$ and $t+dt$, given that detection
occurs at $x$. In this framework, it is shown that energy
measurements of a stationary state display a non-zero dispersion,
and that energy-time uncertainty arises from first principles. We
show that a central result on arrival time, obtained through
approaches that resort to {\it ad hoc} assumptions, is a natural,
built-in part of the formalism presented here.
\end{abstract}
\maketitle


In Schr\"odinger quantum mechanics (QM) there is a clear asymmetry
between time and space. Time is a continuous parameter that can be
chosen with arbitrary precision and used to label the solution
of the wave equation. In contrast, the position of a particle is seen as an
operator, and therefore its value under a measurement is inherently
probabilistic. It is common to hear that this asymmetry is due to
the non-relativistic character of the Schr\"odinger equation (SE).
Although partially correct, this argument is largely insufficient to
justify all the disparity between space and time in the formalism of
QM.

A clear illustration is as follows. In a position measurement,
$\psi(x,t)=\langle x|\psi(t)\rangle$ gives the probability amplitude
of finding the particle within $[x,x + {d}x]$, \emph{given
that} the time of detection is $t$. Would it not be equally reasonable,
even in the non-relativistic domain, to ask about the
probability of measuring the particle between $x$ and $x + {d}x$, and
$t$ and $t + {d}t$? In this broader scenario, inquiring
about the state of a particle at a given time $t$ (as we often do),
should make as much sense as asking about the state of that particle
in a given position $x$ (which we never do). In addition, if
symmetry is to hold at this level, then there should exist a
``mirror'' wave function $\phi(t,x)=\langle t|\phi(x)\rangle$, where
$x$ is a continuous parameter and $t$ is the eigenvalue of an
observable. If the location of particle becomes a physical reality
only when a measurement is made, then it is a tenable position to
expect that time should emerge in the same way. To earnestly
consider these issues is the main goal of this manuscript.

Time has been addressed in different contexts in QM
~\cite{Wigner,Kijowski,Muga,Werner,Hartle,But,review,Grot,Kumar,Gianni,
toperator,time,rovelli,Hal1,Sels,Gala,dias,book1,book2,Hal2,rovelli2,rovelli3,thesis}.
Common to several of these works is the attempt to remain within the
borders of the standard theory. However, the solution to the
arrival-time problem is considered by several authors to lay outside
the framework of QM. It concerns the arrival of a particle in
a spatially localized apparatus, where a time operator may be
defined so that the relation $[{\hat T}, {\hat H}] = i\hbar$ is
satisfied, and the objective is to obtain the probability
distribution for the detection times. This idea gave rise to
numerous studies, e. g., in quantum
tunnelling~\cite{But,review} and lifetime of metastable systems. We will show that the standard
arrival-time distribution, obtained by various approaches that
usually resort to {\it ad hoc} assumptions, is a natural, built-in
part of the formalism presented here.

Recent works~\cite{time,dias} building on a proposal by Page and
Wootters~\cite{Wooters}, present specific similarities to this letter.
As in reference \cite{time}, we also consider a
time variable $t$ with a Hilbert space ${\cal H}_T$ isomorphic to
that of a spinless particle in one dimension, ${\cal H}_X$. However,
our approach does not assume that it is related to an external
clock, or that it is a formal extension of a physical system. In our
formalism ${\cal H}_T$ (intrinsic to the system) is on the same
footing as ${\cal H}_X$. Moreover, differently from
Ref.~\cite{time}, what we propose to be extended is the set of
possible statistical inferences that QM is able to deal with.

Some symmetry between time and position in QM can be found,
although often concealed by the standard
presentation of the theory. An example is the pair of equations:
${\hat H} {\hat U}_t(t,t')=i\hbar (d/dt){\hat U}_t(t,t')$, and
${\hat p}{\hat U}_x(x,x')=i\hbar (d/dx){\hat U}_x(x,x')$, where
${\hat H}$ is the Hamiltonian of the system, ${\hat U} _t$ is the
time evolution operator, $\hat p$ is the momentum operator, and
${\hat U} _x$ is the translation operator. At a formal level, there
is a complete interplay between the pairs $(\hat p, x)$ and $(\hat
H, t)$. In spite of this perfect correspondence at a mathematical
level, there is a physical asymmetry. In the first place, there is
a clear lack of kets $| t \rangle$ satisfying ${\hat H}
|t\rangle\stackrel{\text{?}}{=}(i\hbar \, {d}/{{ d}t})|t\rangle$, by
analogy with ${\hat p}|x\rangle=(i\hbar \, {d}/{{ d}x})\,|x\rangle$.
The existence of such equations, where $x$ and $t$ play formally
similar roles, would only make sense if $\hat{H}$ and $\hat{p}$ were
also considered on the same footing. This does not happen
in standard QM because, while $\hat{p}$ is defined by
its action upon $\psi(x)$, $\hat{H}$ comes from the replacement of
$x$ and $p$ by $\hat{X}$ and $\hat{p}$ in the classical, symmetrized
Hamiltonian $H(x,p)$.

Furthermore, in a formalism intending to promote time to a
physical, observable quantity, the probability of finding a particle
in $[x,x+ {d}x]$ at an infinitely precise time $t$ {\it
must be rigorously zero}. However, the probability density ${\cal P}(x,t)$
of finding the particle in the space and time intervals
$[x, x + {d}x]$ and $[t, t + {d}t]$ is well-defined.
These quantities should be related by
\begin{equation}
\label{prob1}
{\cal P}(x,t)~{d}x {d}t = f(t) |\psi(x,t)|^2 ~{d}x
{d}t\,,
\end{equation}
where $f(t)$ is determined by Bayes rule, according to
which ${\cal P}(x,t)$ is equal to the probability of finding the
particle between $x$ and $x+{d}x$ \emph{given that} the measurement
occurred precisely at $t$, $|\psi(x,t)|^2$, times the probability of
the system being measured between $t$ and $t+{d}t$, $f(t)$, whatever
the outcome. For this reason, we express the wave function with
the more appropriate notation $\psi(x|t)$.

It is essential to realize that the function $f(t)$ cannot be
obtained through the knowledge of $|\psi(t)\rangle$, the solution of
the SE. The temporal weighting function constitutes new information
necessary to express the full state of the system. By ``full state''
we mean the information necessary to predict experimental outcomes
related to statistical inferences other than those with a fixed time, that is, outside the traditional scope of QM.

In addition, due to the symmetry of Bayes rule, we can express the
probability ${\cal P}(x,t)\,{d}x {d}t$ as $|\phi(t|x)|^2 g(x) ~{d}x
{d}t$, where the quantities have analogous roles as those of Eq.
(\ref{prob1}), with $t \rightleftarrows x$. Therefore,
$|\phi(t|x)|^2 dt$ corresponds to the probability of finding the
particle in the time window $[t,t+{d}t]$ \emph{given that} the
position measurement gives exactly $x$, and $g(x)$ is the
distribution associated with position measurements regardless of
$t$. Note that these quantities are not present in standard QM.
Operationally, in order to determine ${\cal P}(x,t)$ one has to fill
the space with detectors and turn them on simultaneously with
respect to the laboratory clock, and wait until one of the
detectors, located at some position $x$, clicks at some time $t$.
This corresponds to an event registered at $(x,t)$. Then, repeat the
procedure several times to extract the statistics described by
${\cal P}(x,t)$.

Knowing that quantum mechanics is a remarkably
successful theory, and considering all the
previous observations, we propose two {\it supplementary} assumptions to
equip QM to deal with a broader set of statistical inferences (corresponding
to valid experimental questions).

{\it Assumption 1}: The minimal Hilbert space necessary
for a complete quantum description of a spinless particle in one
dimension is ${\cal H}={\cal H}_X\otimes {\cal H}_T$, where ${\cal
H}_T$ is as intrinsic to the system as ${\cal H}_X$. Accordingly,
the ket that represents the full state of a quantum particle,
denoted by $|| \Psi\rangle \in {\cal H}$, can be expressed as
\begin{equation}\label{bigpsi}
|| \Psi\rangle=\int \int \Psi(x\, \text{\footnotesize \&} \, t)\, |x
\,t\rangle ~{d}x {d}t\;,
\end{equation}
with $|x\, t\rangle=|x\rangle \otimes  |t\rangle$ and $\langle x'\,
t'||\Psi\rangle= \Psi(x'\, \text{\footnotesize \&} \, t')$. We
interpret $|x\,t\rangle$ as the state of a particle that is observed
at the position $x$ and at the instant $t$. We employed the notation
$\Psi(x\, \text{\footnotesize \&} \, t)$ to make it clear that this
quantity is {\it inequivalent} to the wave function
$\psi(x,t)=\psi(x|t)$. The squared modulus of $\Psi(x\,
\text{\footnotesize \&} \, t)$ is ${\cal P}(x,t)$ in Eq.
(\ref{prob1}). Bayes rule for the amplitudes has the general form
\begin{equation}
\label{bigpsi1}
\Psi(x\, \text{\footnotesize \&} \, t)=\psi(x|t)\sqrt{f(t)}e^{i\alpha(x,t)}={\phi}(t|x)\sqrt{g(x)}e^{i\beta(x,t)}\;,
\end{equation}
with $f(t)\ge 0$ and $g(x)\ge 0$ (we simply set $\alpha=\beta=0$).
For a complete description we need either the first or the second
equality in Eq. (\ref{bigpsi1}), not both. We stress that in an
experiment where the statistics of some observable is done by
selecting a specific time, all the usual results of QM immediately
follow by just knowing $\psi(x|t)$. On the other hand, if the
position of the measurement is fixed as a conditional parameter, as
it happens in arrival-time experiments, one has just to know
$\phi(t|x)$ to predict their results. Finally, whenever the position
$x$ and the time $t$ (both unconstrained) are under measurement, the
results should be given by the wave function $\Psi(x\,
\text{\footnotesize \&} \, t)$. It is worth mentioning that in
previous approaches to the role of time in QM, expressions which are
notationally similar to the space-time integral (\ref{bigpsi}) have
appeared. For example, Eq.~(6) of \cite{rovelli2} has completely
different construction and interpretation, since it corresponds to a
new representation of ordinary states of QM. This state is defined
in order to take into account the time interval of interaction with
the measuring device, and it can be derived, in principle, by using
the SE, as is done in Refs.~\cite{rovelli2,rovelli3}. On the other
hand, Eq.~(\ref{bigpsi}) encompasses a temporal probabilistic
character not obtainable from the traditional formulations.

By writing Eq. (\ref{bigpsi1}) as
$|\psi(x|t)|^2/|{\phi}(t|x)|^2=g(x)/f(t)$ and integrating over $x$
we obtain
\begin{equation}
f(t)=\left[ \int |\psi(x|t)|^2/|{\phi}(t|x)|^2 ~dx\right]^{-1}\;,
\end{equation}
for ${\phi}(t|x)\ne 0$ and $f(t)\ne 0$. This result makes it clear
that the temporal distribution does not depend on the details of
the detectors.

{\it Assumption 2}: By analogy with the standard operators
$\hat{X}$, $\hat{p}$, and $\hat{H}(\hat{X},\hat{p};t)$ acting in
${\cal H}_X$, we define the mirror operators: observation time
$\hat{T}$ (mirror of $\hat X$), Hamiltonian $\hat{h}$ (mirror of
$\hat p$), and momentum $\hat{P}(\hat{T},\hat{h};x)$ (mirror of
$\hat H$) acting in ${\cal H}_T$. The observables in lower cases are
solely defined by their action upon the bases $\{|x\rangle\}$ and
$\{|t\rangle\}$, through the relations
\begin{equation}
\label{h} \frac{\langle
xt|\hat{p}||\Psi\rangle}{i\hbar}=-\frac{\partial}{\partial x}
\Psi(x\, \text{\footnotesize \&} \, t),~\frac{\langle
xt|\hat{h}||\Psi\rangle}{i\hbar}=\frac{\partial}{\partial t}
\Psi(x\, \text{\footnotesize \&} \, t).
\end{equation}
These operators are canonically conjugated to $\hat{X}$ and
$\hat{T}$, respectively, which are defined by
$\hat{X}|xt\rangle=x|xt\rangle$ and ${\hat{T}}
|xt\rangle=t|xt\rangle$.
The inverse
Fourier transform of $|x\,t\rangle$ reads
$|x\,t\rangle=1/(2\pi\hbar)\int \int \exp(-ipx/\hbar+i\varepsilon
t/\hbar)~|p\,\varepsilon\rangle ~{d}p {d}\varepsilon$ with
$|p\,\varepsilon\rangle=|p\rangle \otimes |\varepsilon\rangle$,
$|p\rangle \in {\cal H}_X$ and $|\varepsilon\rangle \in {\cal H}_T$.
This implies $|| \Psi\rangle=\int \int \tilde{\Psi}(p\,
\text{\footnotesize \&} \, \varepsilon)\, |p \,\varepsilon\rangle
~{d}p {d}\varepsilon$, where
$\tilde{\Psi}(p\, \text{\footnotesize \&} \,
\varepsilon)=\frac{1}{2\pi\hbar}\int \int
\exp[-i(px-\varepsilon t)/\hbar] \Psi(x\, \text{\footnotesize
\&} \, t) ~{d}x {d}t$.

The expectation value of energy, e.g., is given by an
average over the whole Hilbert space ${\cal H}={\cal H}_X\otimes
{\cal H}_T$:
\begin{equation}
\label{meanenergy} \langle {h} \rangle= \int \int |\tilde{\Psi}(p\,
\text{\footnotesize \&} \, \varepsilon)|^2\,\varepsilon\,{d}p
{d}\varepsilon\;,
\end{equation}
an analogous relation holding for the linear momentum. Note that
these averages are given by the lower case observables.
Definition~(\ref{meanenergy}) leaves mean values of energy unchanged
while variances may change with respect to the standard theory. This
is not in contradiction with QM since Eq.~(\ref{meanenergy}) is
defined in a different way from the mean value of traditional QM,
where time is fixed. Here, we take into account an intrinsic
probabilistic character of the detection moment and, consequently,
an extra integration over time is necessary [see
Eq.~(\ref{bigpsi})]. A compelling consequence of these relations is
that, whenever a system is under measurement and it is not
possible/desirable to fix the time, there is a nonzero variance
associated with its energy, even if its state is stationary, as we
will verify in what follows.
%
%

Consider the general problem of a confined particle with a
Schr\"odinger state being:
$|\psi(t)\rangle=\exp(-iE_nt/\hbar)~|\psi_n\rangle$, where $\hat{H}|\psi_n\rangle=E_n|\psi_n\rangle$.
To derive $f(t)$, suppose that the particle can be detected all
over the region where the wave function is non-vanishing.
If the particle has a finite probability $q$ of being measured after
$t=0$ and before $t=\delta t$, but it turns out that the detection
did not happen, for a Markov process, the probability of it occurring between
$t=\delta t$ and $t=2\delta t$ is the same as it was in $t=0$. Thus, the probability
of no observation up to $t=n\delta t$ is $P(t=n\delta
t)=(1-q)^n$. If $\delta t$ is sufficiently small, we can assume $q
\ll 1$, so that $P(t=n\delta t)\simeq {\rm e}^{-n\delta t
\Lambda}={\rm e}^{-\Lambda t}$, where we defined $\Lambda\equiv
q/\delta t$. The associated probability density is $f(t)=-dP/dt$,
leading to the Poisson distribution $f(t)=\Lambda{\rm e}^{-\Lambda
t}$. Combining Eq. (\ref{bigpsi1}) and the previous result, the
complete state of the system can be written as
\begin{equation}
\label{Psi0}
||\Psi_n\rangle
=|\psi_n\rangle\otimes\int  \sqrt{\Lambda}~{\rm e}^{\left(-\Lambda
t/2 -iE_nt/\hbar \right)}~
|t\rangle~
dt\;.
\end{equation}
The expectation time associated with the occurrence of the
observation is easily obtained and reads $\langle T \rangle=
1/\Lambda$, as it should do. The time uncertainty is ${\Delta T}
=\sqrt{\langle T^2 \rangle - \langle T \rangle^2}= 1/\Lambda$.
Complementarily, we have
\begin{eqnarray}
\label{Phii} \tilde{\Psi}(p\, \text{\footnotesize \&}\,
\varepsilon)&=&\sqrt{\Lambda} \int~ \tilde{\psi}_n(p) ~{\rm
e}^{i[(\varepsilon-E_n)/\hbar+i\Lambda/2] t} ~dt,
\end{eqnarray}
leading to the energy-momentum probability density given by
$|\tilde{\Psi}(p\, \text{\footnotesize \&}\, \varepsilon)|^2
=|\tilde{\psi}_n(p)|^2~|\chi(\varepsilon)|^2$, with
\begin{eqnarray}
\label{e}
|\chi(\varepsilon)|^2=\frac{1}{\pi}\,\frac{\hbar\Lambda/2}{(\varepsilon-E_n)^2+(\hbar\Lambda/2)^2}\;.
\end{eqnarray}
Note that $\tilde{\Psi}(p\, \text{\footnotesize \&}\, \varepsilon)$
is factorable because the initial wave function is a stationary
state \cite{comment2}. By replacing $|\tilde{\Psi}(p\,
\text{\footnotesize \&}\, \varepsilon)|^2$ into
Eq.~(\ref{meanenergy}), we have $\langle { h} \rangle  = \int
|\chi(\varepsilon)|^2\varepsilon~ d\varepsilon$. Thus, the result of
an energy measurement is $\varepsilon$, satisfying the Lorentzian
distribution (\ref{e}).

The distribution $|\chi(\varepsilon)|^2$ does not have a
well-defined variance due to its fat tails. However, its full width
at half maximum, $\delta \varepsilon$, is $\hbar\Lambda$. Thus,
${\Delta T}{\delta \varepsilon} \sim 1/\Lambda \times \hbar\Lambda=
\hbar$. Result (\ref{e}) predicts an energy linewidth similar to the
natural linewidth which arises from the interaction with the
electromagnetic vacuum. In both cases the profile is Lorentzian, but
the physical origins are completely different, since, in our
formalism the linewidth appears because the detection time is
considered as a probabilistic variable. Since linewidth measurements
do not constrain the observation time, our formalism should apply
and Eq.~(\ref{Phii}) would give the actual profile as a convolution
of the Lorentzian in (\ref{e}) with that describing the natural
linewidth. The result is also a Lorentzian, but broader (width
$\Lambda+\Gamma$) and with a lower peak [height
$1/\pi(\Lambda+\Gamma)]$, where $\Gamma$ is the spontaneous decay
rate. Therefore, if one is able to measure the natural linewidth
minimizing all other broadening effects (Doppler effect, collisional
effect, etc), then, the measured width should be {\it larger} than
that predicted by QM (e. g., via {\it ab initio} calculations).
Because $\Lambda$ may be small, we may need to address situations
for which $\Gamma$  and $\Lambda$ do not differ by more than a few
orders of magnitude. Thus, it would be easier to observe this
possibly subtle difference, if it exists, in long-lived systems
(narrow linewidths).
%
%

We now derive the dynamic equation for $\phi(t|x)$, ``dynamic''
meaning how $\phi$ changes with $x$. We will do it through a direct
analogy with SE. Let us define $|\phi(x)\rangle \in {\cal H}_T$,
where $x$ is a parameter that can be chosen arbitrarily in the same
way as the time $t$ in the standard theory. In addition, the ket
$|t\rangle$ corresponds to the relative state of a particle that is
observed at time $t$, so that $\langle t|\phi(x)\rangle= \phi(t|x)$.

Due to the isomorphism between ${\cal H}_X$ and ${\cal H}_T$,
we must have $\langle t|t'\rangle=\delta (t-t')~~~{\rm
and}~~~\mathbb{I}=\int |t\rangle\langle t|~dt$, the
orthogonality of $\{|t\rangle\}$ ensuring that the particle is observed
at a specific time. With these definitions, we can write
$|\phi(x)\rangle = \int \phi(t|x)~ |t\rangle ~dt$ similarly to
$|\psi(x)\rangle = \int \psi(x|t)~ |x\rangle ~dx$.
To make sure that the particle will be observed
during the measurement process, the state has to be normalized,
$\langle\phi(x)|\phi(x)\rangle=1$, which implies $\int |\phi(t|x)|^2
dt=1$. Finally, we interpret $|\phi(t|x)|^2 dt$ as the probability
of measuring the particle in the time interval $[t,t+dt]$, given
that it is observed at the position $x$.

We proceed by attributing to the momentum operator $\hat P$ acting in ${\cal
H}_T$ the same status and role as $\hat H$ in conventional QM.
 Also, we use Eq. (\ref{h}) to write $\langle t'| {\hat h}|t
\rangle = \delta(t-t')\left(i\hbar d/dt \right)$, which
automatically leads to the canonical commutation relation $[\hat
h,\hat T]=i\hbar$. In addition, recall that SE describes how
$|\psi(t)\rangle$ changes under ``time translations'': $\label{sch}
{\hat H} |\psi(t)\rangle = i\hbar (d/dt)|\psi(t)\rangle$, with
${\hat H}={\hat p}^2/(2m) + {\hat V(\hat X,t)}$. The analogous
relation for kets in ${\cal H}_T$ is the space-dependent SE:
\begin{equation}
{\hat P}|\phi(x)\rangle = i\hbar \frac{d}{dx} |\phi(x)\rangle,
{\hat P}=\pm \sqrt{ 2m\left[ {\hat h} - {\hat V(x,{\hat T})}\right]}.
\end{equation}
Since $\hat P$ has two branches (signs $\pm$), we assume
that $\phi(t|x)$ is a two-component pseudospinor:
\begin{equation}
\label{phig}\phi(t|x)=\left( \begin{array}{c} \phi^+(t|x) \\
\phi^-(t|x) \end{array} \right),
\end{equation}
and the mirror equation is in fact
\begin{equation}
\label{eqphin} \hat{\sigma}_z \sqrt{2m \left[{ i\hbar \frac{d}{dt}} -
V(x,t)\right]}~ \phi(t|x) = i\hbar \frac{d}{dx}\phi(t|x),
\end{equation}
where $\hat{\sigma}_z=$diag$(+1,-1)$. It is then clear that
$\sqrt{g}$ in Eq. (\ref{bigpsi1}) is to be understood as a vector
with components $\sqrt{g}^{\pm}$. Equation~(\ref{eqphin}), one of
our central results, is a Schr\"odinger-like equation for
$\phi(t|x)$, where the position $x$ of the observation is a
conditional parameter. Finally, we define in the usual way the
associated probability density as $\rho= |\phi(t|x)|^2 \equiv
\phi^{\dagger}(t|x) \phi(t|x)$.

Pauli pointed out the impossibility of defining a self-adjoint time
operator conjugated to a Hamiltonian with a spectrum bounded from
below~\cite{Pauli}. Pauli's result is a no-go theorem constraining
the possible time observables derived by using standard quantum
theory. The suggested framework does not suffer from this limitation
since it is clearly not contained in traditional QM.
Moreover, it is not necessary for $\hat h$ to be bounded from below
since it plays the same role as $\hat p$ in standard quantum theory.
Positive values of energy may be required as a consequence of the
boundary conditions on $\phi(t|x)$. In these circumstances, the
commutation relation $[\hat h,\hat T]=i\hbar$ automatically leads to
the energy-time uncertainty $\Delta \varepsilon \Delta T \ge
\hbar/2$, where $\Delta \varepsilon$ and $\Delta T$ are the
root-mean-square deviations of $\hat h$ and $\hat T$, which act in
${\cal H}_T$. Note that, here, it is the customary relation $[\hat P,\hat x]=i\hbar$
that cannot be derived, since $x$ is a parameter.

Hereafter, we focus on the wave equation for the free particle
[$V(x,t)=0$]. By inspecting Eq.~(\ref{eqphin}), we can obtain the
temporal eigenfunction for the momentum operator defined as ${\hat
P}\phi_P(t)=P\phi_P(t)$, where ${\hat P}=\hat {\sigma}_z
\sqrt{2m\left(i\hbar d/dt\right)}$. It is worth noting the analogy
between the eigenfunction $\phi_P(t)$ (a space-independent function)
and the time-independent Schr\"odinger state $\psi_E(x)$, which
satisfies ${\hat H}\psi_E(x)=E \psi_E(x)$. In this latter case, the
eigenenergy solution is simply $ \psi_E(x|t)=
\psi_E(x){\exp}(-iEt/\hbar)$. Accordingly, we write $\phi(t|x)
\equiv \phi_P(t|x) =\phi_P(t)~{\exp}(iPx/\hbar)$. By substituting
the previous definition into Eq.~(\ref{eqphin}), we have
$\hat{\sigma}_z \sqrt{2m(i\hbar d/dt)}~\phi_P(t) = P\phi_P(t)$. The
last step is to
identify $\sqrt{d/dt}$ with the Riemann-Liouville fractional
derivative $_{-\infty}D^{1/2}_t$, which is equivalent to the Caputo
fractional derivative~\cite{frac}. This leads to
\begin{equation}
\label{eqphi6}
\hat{\sigma}_z \sqrt{2m i \hbar}~_{-\infty}D^{1/2}_t
~\phi_P(t) = P ~\phi_P(t).
\end{equation}
Let us consider a solution such as
$\phi^{\pm}_P(t)=C^{\pm}_P \exp(-iwt)$, and use the identity
$_{-\infty}D^{1/2}_t \exp(-iwt)=\sqrt{-iw} \exp(-iwt)$. By
doing this, we readily obtain the dispersion relation
$P=\pm \sqrt{2m\hbar w}$, where we did not use negative
energies since they lead to imaginary $P$.

Because Eq.~(\ref{eqphin}) is linear, the general solution
is $\phi^{\pm}(t|x)=\int_0^{\infty}
A^{\pm}_P~{\exp}(-iE_Pt/\hbar)~{\exp}(\pm iPx/\hbar)~dP$, $E_P
\equiv \hbar w=P^2/2m$. By setting $A^{\pm}_P\equiv
C^{\pm}_P~\sqrt{|P|/2\pi m}$, the temporal normalization condition
for $\rho$ reads $\int_{-\infty}^{\infty}\rho(t|x)~dt=1$
$\Rightarrow$ $\int_{-\infty}^{\infty}
\left(|C^+_P|^2~+~|C^-_P|^2\right)dp=1$, where $|C^{\pm}_P|^2$
corresponds to the probability density of finding the particle with
momentum $\pm P$. With this, we can express the general solution as
\begin{equation} \label{gnf}
\nonumber
\phi(t|x)=\frac{1}{\sqrt{2\pi m\hbar}}\int_0^{\infty} ~ \left(
\begin{array}{c}
C^+_P~\sqrt{P}~{\rm e}^{iPx/\hbar}\\
C^-_P~\sqrt{P}~{\rm e}^{ -iPx/\hbar} \end{array} \right) ~ {\rm
e}^{-iE_Pt/\hbar}~dP.
\end{equation}
It is natural to define the states $|P^\pm\rangle$ so that $\langle
t|P^\pm\rangle = \phi^{\pm}_P(t)\equiv \sqrt{P/2\pi
m\hbar}~\exp(-iE_Pt/\hbar)$. In this way, we can write
$\phi^{\pm}(t|x)=\int C^{\pm}_P~\phi^{\pm}_P(t)~\exp(\pm
iPx/\hbar)~dP$. These states are the same eigenstates (with positive
and negative momentum) as the time-of-arrival operator defined by
adding the symmetrization and quantization of the classical
expression $mx_{class}/p_{class}$ to conventional QM.
This operator was first defined by Aharonov and Bohm~ \cite{Aharo},
and later used by several other authors (see, for instance,
~\cite{Muga,Gianni,Goto}).

Note the symmetry between the general solutions for $\psi(x|t)$ and
$\phi(t|x)$: $\phi^{\pm}(t|x)=\int C^{\pm}_p\phi^{\pm}_p(t)\exp(\pm
iPx/\hbar)dp$ and $\psi(x|t)=\int C_E\psi_E(x)
{\exp}(-iEt/\hbar)dE$. The function $\phi(t|x)$ is a
superposition of momentum eigenstates, whereas $\psi(x|t)$
corresponds to a linear combination of states with well-defined
energy. With this analogy, we can interpret $|\phi_P(t)|^2$ as
the probability density of observing the particle at the instant
$t$, given that it has momentum $P$, while $|\psi_E(x)|^2$ is the
probability density of finding the particle at the position $x$,
given that the energy is $E$. A fundamental issue related to time in
QM is to seek a temporal distribution that describes
the instant of time at which a certain property of a system assumes
a given value~\cite{DelMuga}. This is exactly the interpretation of
$|\phi_P(t)|^2$ with the momentum $P$ as the physical property.

Finally, the probability density of finding the
particle at the instant $t$, given that a measurement is performed
at the position $x$, is given by
\begin{eqnarray} \label{pd}
\rho(t|x)&=& \frac{1}{2\pi m\hbar}
\Bigg\{~{\Bigg|}\int_0^{\infty}~C^+_P~\sqrt{P}~{\rm
e}^{iPx/\hbar-iE_P t/\hbar}~dP{\Bigg |}^2
\nonumber\\
&+& {\Bigg |} \int_0^{\infty}~C^-_P~\sqrt{P}~{\rm e}^{-iPx/\hbar -
iE_P t/\hbar}~dP{\Bigg |}^2 ~\Bigg\}.
\end{eqnarray}
This is exactly the time-of-arrival probability density
obtained by several authors via different
approaches~\cite{Grot,All,Kijo,DelMuga,Baute} and is in excellent
agreement with numerical ``quantum jump'' time-of-flight
simulations~\cite{book1}. However, the models employed have faced
problems even in the free particle case. In
Refs.~\cite{Kijo,DelMuga}, e. g., Eq. (\ref{pd}) was obtained from
the Schr\"odinger current density, which is not positive definite.
It has been argued that there were \emph{ad hoc} assumptions not
included in standard quantum theory~\cite{Miel}, e. g., the
association of the signs in $\pm P$ with the direction of arrival
\cite{Leavens}. Moreover, the time operator put forward by Aharonov
and Bohm is semiclassical and system-dependent, since it is obtained
by the quantization of the classical time of arrival at a certain
point $x$. Here, on the other hand, we develop a wave dynamics where
temporal probability densities arise from first principles [the
dynamic equation~(\ref{eqphin})] through a positive definite
quantity $\rho$, similarly to $|\psi(x|t)|^2$ in standard QM.
The distribution $\rho$ is obtained with no
dependence on the properties of a particular measuring apparatus.


We argue that part of the asymmetry between position and time in
non-relativistic quantum mechanics is due to the fact that we
experience these degrees of freedom in drastically different ways,
and not only because of the lack of relativistic covariance.
We tend to face time as a parameter much more naturally than position,
although this inclination is not justifiable, on logical grounds.
Guided
by the requirement of symmetry between $x$ and $t$ as statistical
variables and by Bayes
theorem, we find a mirror wave function that gives new physical
information, not obtainable through the knowledge of the
Schr\"odinger wave function. We find the corresponding equation of
motion, and show that the arrival-time distribution follows
naturally. In contrast, previous derivations resort to assumptions
which demand that one either gives up on the hermiticity of the
operator $\hat T$ or on the validity of its canonical commutation
relation with the Hamiltonian~\cite{DelMuga}. Here we kept both
desirable properties, and found that actual measurements on an
energy eigenstate lead to results with a non-zero dispersion,
which illustrates how the energy-time
uncertainty arises in the formalism.

Note the nature of the supplementation we propose. In situations where time is,
in any way, fixed, standard QM emerges unchanged. However, QM does not provide any
obvious answer to valid experimental questions related to other kinds of inference. We believe
that the reason is made clear in the present work, which also provides a plausible fill to this gap.

\begin{acknowledgments}
E. O. D. acknowledges financial support from FACEPE through its PPP
Project No. APQ-0800-1.05/14. F. P. thanks financial support from
CNPq through the Instituto Nacional de Ci\^encia e
Tecnologia~-~Informa\c{c}\~ao Qu\^antica (INCT-IQ).
\end{acknowledgments}


\begin{thebibliography}{99}
\bibitem{comment2} Had we considered a coherent superposition, the energy-momentum distribution
would be correlated.
\end{thebibliography}
\end{document}